%% file: sample-acmsmall.tex
\documentclass[10pt,a4paper]{article}
\usepackage{geometry}

\usepackage[utf8]{inputenc}

\usepackage{cite}

\usepackage{balance}       % to better equalize the last page
\usepackage{graphics}      % for EPS, load graphicx instead 
\usepackage[T1]{fontenc}   % for umlauts and other diaeresis
\usepackage{txfonts}
\usepackage{mathptmx}
\usepackage[pdflang={en-US},pdftex]{hyperref}
\usepackage{color}
\usepackage{tabularx}
\usepackage{multirow}
\usepackage{booktabs}
\usepackage{textcomp}
\usepackage{siunitx}
\usepackage{array}
\usepackage[gen]{eurosym}
\usepackage[dvipsnames, table]{xcolor}
\usepackage{cite} 
\usepackage{booktabs}
\usepackage{tabularx}
\usepackage{xcolor}
\usepackage{textcomp}
\usepackage{colortbl}
\usepackage{eurosym}
\usepackage{multirow}
\usepackage[T1]{fontenc}
\usepackage{siunitx}
\usepackage{url}
\usepackage{adjustbox}
\usepackage{microtype}        % Improved Tracking and Kerning
\usepackage{ccicons}          % Cite your images correctly!
\usepackage{todonotes}

\usepackage{xspace}

\usepackage{mdframed}

\definecolor{rowcol}{rgb}{0.9,0.9,0.9}
\newcounter{methodEnumCounter}

\usepackage{url}

\usepackage{microtype}
\DisableLigatures[f]{encoding = *, family = * }

\usepackage[aboveskip=1pt,labelfont=bf,labelsep=period,singlelinecheck=off]{caption}

\makeatletter
\renewcommand{\@biblabel}[1]{\quad#1.}
\makeatother

\usepackage{lastpage,fancyhdr,graphicx}
\usepackage{epstopdf}
\fancyheadoffset[L]{2.25in}
\fancyfootoffset[L]{2.25in}

\usepackage{color}

\definecolor{Gray}{gray}{.25}

\usepackage{graphicx}

\usepackage{sidecap}

\usepackage{wrapfig}
\usepackage[pscoord]{eso-pic}
\usepackage[fulladjust]{marginnote}
\reversemarginpar

\definecolor{rowcol}{rgb}{0.9,0.9,0.9}

\begin{document}
\vspace*{0.35in}

% title goes here:
\begin{flushleft}
{\Large
\textbf{\newline{A Method and Analysis to Elicit User-reported Problems in Intelligent Everyday Applications}}
}
\newline
% authors go here:
\\
Malin Eiband\textsuperscript{1},
Sarah Theres V\"{o}lkel\textsuperscript{1},
Daniel Buschek\textsuperscript{2},
Sophia Cook\textsuperscript{1}
Heinrich Hussmann\textsuperscript{1}
\\
\bigskip
{1} LMU Munich
\\
{2} University of Bayreuth
\\
%bigskip
%* malin.eiband@ifi.lmu.de

\end{flushleft}

%\section*{Abstract}
\input{sections/00-abstract.tex}

% the * after section prevents numbering

\input{sections/01-introduction.tex}
\input{sections/02-related-work.tex}
\input{sections/03-method.tex}
\input{sections/04-results-topics.tex}

\input{sections/05-online-survey.tex}
\input{sections/06-results-survey.tex}

\input{sections/07-discussion.tex}

\input{sections/08-conclusion.tex}

%This is where your bibliography is generated. Make sure that your .bib file is actually called library.bib
\bibliography{bibliography}

%This defines the bibliographies style. Search online for a list of available styles.
\bibliographystyle{abbrv}

\end{document}

%% file: sections/00-abstract.tex
\begin{abstract}
The complex nature of intelligent systems motivates work on supporting users during interaction, for example through explanations. However, as of yet, there is little empirical evidence in regard to specific problems users face when applying such systems in everyday situations.
This paper contributes a novel method and analysis to investigate such problems \textit{as reported by users:}
We analysed 45,448 reviews of four apps on the Google Play Store (Facebook, Netflix, Google Maps and Google Assistant) with sentiment analysis and topic modelling to reveal problems during interaction that can be attributed to the apps' algorithmic decision-making. We enriched this data with users' coping and support strategies through a follow-up online survey (N=286). In particular, we found problems and strategies related to content, algorithm, user choice, and feedback.
We discuss corresponding implications for designing user support, highlighting the importance of user control and explanations of output, rather than processes.

Author pre-print, to appear in \textit{Special Issue on Highlights of ACM Intelligent User Interface (IUI) 2019. ACM Trans. Interact. Intell. Syst.}

\end{abstract}

%% file: sections/01-introduction.tex
\section{Introduction}
Algorithmic decision-making has permeated many interactive systems that people use on a daily basis (e.g. film recommendations, social networks, navigation). Based on Jameson and Riedl~\cite{jameson2011}, we define that an intelligent system ``embodies one or more capabilities that have traditionally been associated more strongly with humans than with computers, such as the abilities to perceive, interpret, learn, use language, reason, plan, and decide''. Hence, intelligent systems select what kind of information is to be considered relevant, influence which content users do or do not see, draw inferences from user behaviour, and shape future behaviour~\cite{Gillespie2014}. 

This algorithmic decision-making poses difficult challenges for human-computer interaction (HCI) since intelligent systems violate established usability principles, such as easy error correction and predictable output~\cite{Amershi2014, Dudley2018}.
As a result, HCI and related fields have recognised the need to support users during interaction with intelligent systems~\cite{ACM2017, Diakopoulos2016, Hager2017}.
Potential problems that users may face in interaction with such systems have either been examined on a general level or tied to evaluations of specific prototypes:

On the general level, for example, people may lack awareness of algorithmic decision-making in consumer applications~\cite{Eslami2015}. Moreover, uncertainty and lack of knowledge about such decision-making may cause what has been called ``algorithmic anxiety''~\cite{Jhaver2018}. Related, researchers have also observed ``algorithmic aversion''~\cite{Dietvorst2014}, when users put less confidence in algorithmic than human predictions, even if the latter are less accurate. 

On the other end, possible solutions for supporting users in interactions with intelligent systems have been realised as specific research prototypes, for example, in work on explanations~\cite{Kulesza2013}, scrutability~\cite{Kay2013}, interactive machine learning~\cite{Dudley2018}, and end-user debugging~\cite{Kulesza2015}.

We argue that the actual meeting point of humans and intelligent systems \textit{in practice} today can be located somewhere in between these general and prototype levels -- in publicly available intelligent everyday applications. In this paper we focus on \textit{consumer} applications, which end-users employ on a daily basis, such as social networks, media service providers, and voice assistants.  
However, as of yet, there is little empirical evidence on practical problems with such intelligent everyday applications. The research community thus risks that its concepts and prototype solutions remain decoupled from everyday problems and user needs. Moreover, unawareness of current practical problems may hinder addressing relevant problems in future research. 

In this paper, we aim to assess user problems in intelligent everyday applications, as well as users' coping strategies and desire for support. In particular, we investigate the following questions:
\begin{enumerate}
    \itemsep-0.1em
    \item \textit{Which problems do users encounter when using intelligent everyday applications?}
    \item \textit{What kind of support do users want for which problem?}   
\end{enumerate}

To address these questions, we contribute a novel method and analysis to investigate problems that can be attributed to algorithmic decision-making \textit{as reported by users}. We base our analyses on app reviews of four apps on the Google Play Store that are used in everyday contexts, namely Facebook, Netflix, Google Maps, and Google Assistant. Using sentiment analysis and topic modelling, we reveal problems which we then investigate in depth in a follow-up online survey (N=286). In the survey, we describe the extracted problems as scenarios to collect participants' experiences in these situations and to assess their coping strategies and need for support.

In sum, our contribution in this article is as follows:
(1) We present and reflect on an method to capture user problems with intelligent everyday applications that combines both automatic and manual analysis based on user reviews. (2) We provide empirical evidence for such problems as reported by users. (3) We extract key problems as well as users' strategies to cope with these problems and their wishes for support. (4) We present and discuss implications for supporting users in intelligent everyday applications.

Our work thus connects and structures practical problems and research directions as well as solutions. We hope to facilitate better understanding of problems users face when interacting with intelligent systems in everyday contexts. 

%% file: sections/02-related-work.tex
\section{Background and Related Work}
We focus on intelligent systems that mediate \textit{everyday} tasks and practices~\cite{Willson2017}.

\subsection{Assessing User Problems with Intelligent Systems}
The impact of such systems has received considerable attention in the last years. The application context of social media in particular has sparked interest, both among the general public and HCI researchers. 
For example, Bucher~\cite{Bucher2017} examined and catalogued situations in which users become aware of and experience algorithmic decision-making on Facebook, such as when they note being profiled or ``found'' by the algorithm. As an example, users mentioned that they were drinking a cup of coffee while seeing an ad for the same coffee brand in the feed.

Eslami et al.~\cite{Eslami2015} found that users' awareness of the Facebook news feed curation algorithm was generally low: More than half of their participants (\SI{62.5}{\%}) did not know that the content they see on Facebook is algorithmically selected, but had rather assumed that friends and family were actively hiding posts from them. In another study the year after~\cite{Eslami2016}, Eslami et al. assessed user beliefs about Facebook's curation algorithm and how such beliefs affect their behaviour. A similar study was conducted by Rader and Gray~\cite{Rader2015}, who classified a wide range of user beliefs about Facebook's news feed algorithm.

Other specific application areas in which user-reported problems have been assessed are health and well-being~\cite{Eiband2018a} as well as sharing economy platforms~\cite{Jhaver2018}: The work by Eiband et al.~\cite{Eiband2018a} addressed the problem that users erroneously thought that a human, and not an algorithm, assembled their fitness workouts. Jhaver et al.~\cite{Jhaver2018} looked into the effects of algorithmic evaluation on AirBnB hosts. They found that uncertainty about the workings of the algorithm and perceived lack of control caused what the researchers called ``algorithmic anxiety''.  
        
\subsection{Addressing User Problems in Intelligent Systems}
Given the problems and related negative experiences users may encounter in interaction with intelligent systems, supporting users in such situations has been recognised as a crucial challenge for HCI and related fields~\cite{ACM2017, Diakopoulos2016}.

A particular concept which has received attention in this context is \textit{seamfulness}~\cite{Abdul2018, Alvarado2018, Hamilton2014}. Seamful (in contrast to seamless) techno\-logy design makes system properties apparent to users~\cite{Chalmers2003}. In terms of intelligent systems, a seamful approach might expose (part of) the algorithmic decision-making at work in the interface~\cite{Hamilton2014}.
For example, Eslami et al.~\cite{Eslami2015} developed a prototype to illustrate the effect of algorithmic decision-making on the content users get to see. Another solution has been presented by Munson et al.~\cite{Munson2010}, who built a tool to make users aware of possible biases introduced by the algorithm.

Another substantial body of work focuses on helping users make sense of intelligent systems, for example through explanations.
Such explanations often target the algorithmic decision-making process or a particular output instance~\cite{Rader2018} and their potential benefits have been investigated in diverse studies (e.g.~\cite{Cramer2008, Cramer2009, Kizilcec2016, Kulesza2013, Lim2011, Pu2006, Rader2018}, also cf.~\cite{Abdul2018}). 

A third related line of research investigates interactivity of machine learning: Here, users and their feedback and corrections to the system are an integral part of the machine learning process (see~\cite{Dudley2018} for a review of the field). 
Related, Sarkar~\cite{Sarkar2015} introduces the term ``metamodels'' to describe models about machine learning models as structure for interaction. In particular, these metamodels seek to facilitate user understanding and assessment of the correctness of the system's workings.

As an overarching view on interaction with intelligent systems, Alvardo et al.~\cite{Alvarado2018} propose the concept of ``algorithmic experience''. This concept includes fostering awareness of algorithmic decision-making and how it works as well as deliberately activating or deactivating algorithmic influence. 

\subsection{Linking Problems and Solutions}
As shown in the previous sections, several lines of research seek to assess and address problems users have with intelligent systems and algorithms.
However, to the best of our knowledge, a large-scale empirical assessment of current everyday problems with intelligent systems is still missing. Moreover, we see the need to consider possible solutions for support \textit{along} with the identified problems since these solutions may be specific to particular situations or contexts. For example, Bunt et al.~\cite{Bunt2012} found that the perceived cost of reading explanations in everyday applications tended to outweigh their benefits.
Finding problems and solutions as well as a link between them motivates our research presented in this paper.

%% file: sections/03-method.tex
\section{Analysis Scope}
We focused on four everyday apps which incorporate intelligent algorithmic decision-making: \emph{Facebook}, \emph{Netflix}, \emph{Google Maps} and \emph{Google Assistant}. Social media, recommenders for entertainment, navigation and voice assistants are investigated and discussed in the literature as interesting application domains for interaction with intelligent systems (e.g.~\cite{Alvarado2018, Eslami2016, Ricci2015, Willson2017, beneteau2019}). We found Facebook, Netflix and Google Maps to be the most popular representatives of these domains as per their number of downloads in the Google Play Store in July 2018. Google Assistant and Apple's Siri were the most popular voice assistants in 2019, yet Siri is not available in the Google \mbox{Play Store~\cite{microsoft2019}.}

\emph{Facebook} hosts one of the biggest social networks on the web with more than 2.23 billion active users in 2018~\cite{facebook2018}. Facebook users are presented with a news feed that is algorithmically curated by taking into account user behaviour such as \emph{likes}~\cite{Eslami2015}. Facebook does not disclose its algorithm, which resulted in the formation of folk theories about its working among users~\cite{Eslami2016}.

\emph{Netflix} is a recommender system for streaming films and TV shows. To facilitate users' choice of a film or TV show, Netflix uses a variety of supervised and unsupervised machine learning techniques for personalised recommendations~\cite{gomez-uribe2016}.

\emph{Google Maps} is the most popular navigation app which provides real-time GPS navigation, traffic and transit information as well as points of interests such as restaurants and events. Based on Dijkstra's shortest path algorithm, Google Maps uses deep learning and supervised machine learning techniques to suggest routes~\cite{ibarz2017}.

\emph{Google Assistant} is Google's digital voice-commanded assistant. It uses natural language processing to allow for speech interaction. Google Assistant can search for information online, schedule events and alarms, communicate with other devices and services, or adjust hardware settings, among other functions~\cite{googleassistant2019}.

\section{Research Method}
\label{sec:method}
Investigating problems with intelligent everyday applications is challenging. Terminology like \emph{algorithm} or \emph{intelligent system} is often not part of a layperson's active vocabulary. Simply asking users about their experience with such applications is therefore unlikely to yield fruitful answers. 

To overcome this obstacle, we developed a new research method: It first assesses real user problems by analysing existing sources of reported experiences that users had with intelligent systems. Second, the identified problems are presented to a wider group of users to ask for reflections and desirable solutions. The results can then inform broader solution principles. At the same time, this second part serves as a validation for the extracted problems.

\begin{figure}[t]
  \centering
  \includegraphics[width=\columnwidth]{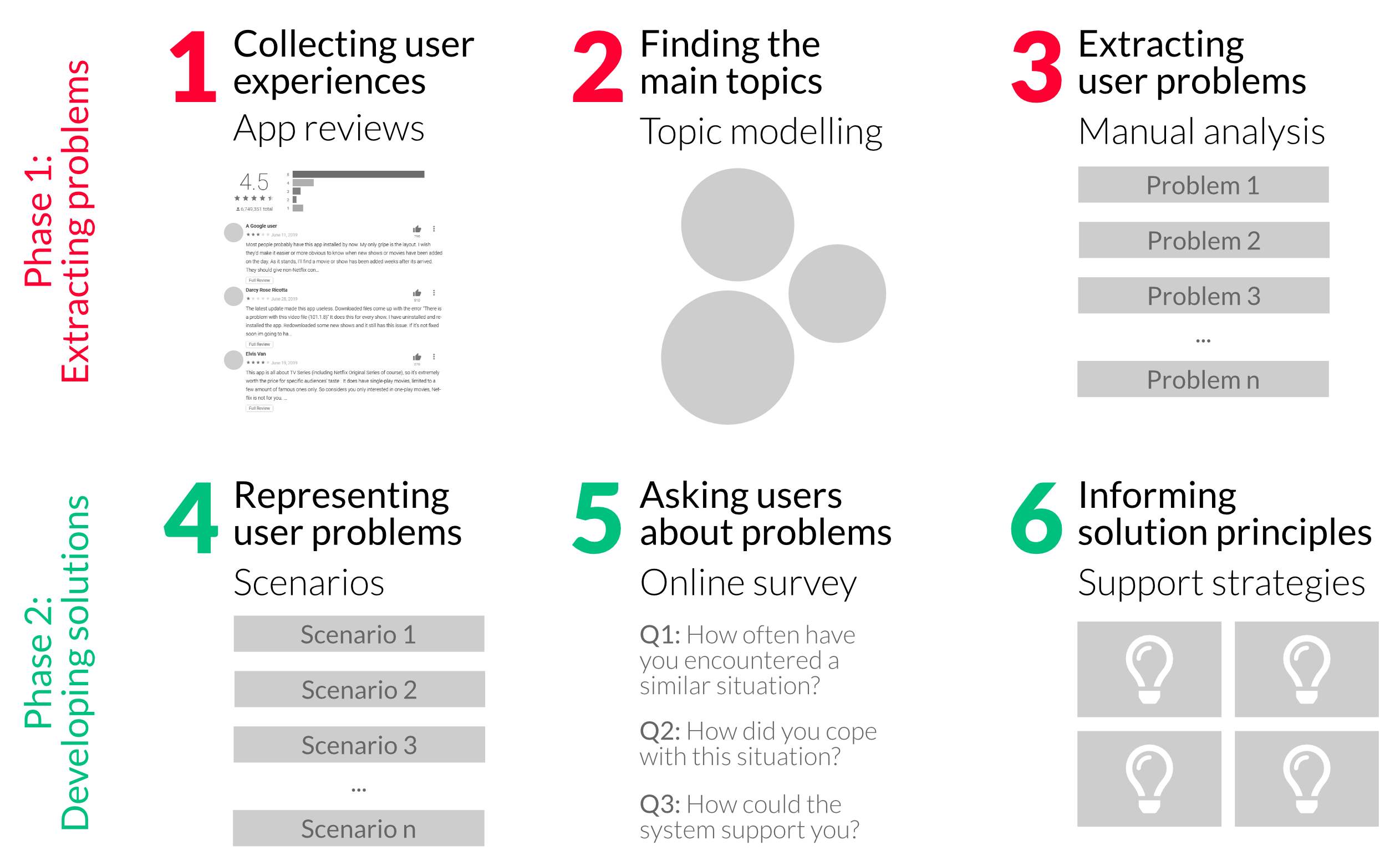}
  \caption{Overview of our research method: It involves two larger phases (rows) with three steps each: The first phase extracts concrete problems that users face in everyday interactions with intelligent systems. To achieve this, we analyse existing sources of reported experiences -- app reviews in our study. The second phase develops solution principles for these problems, while again considering the views of real users, in our study via scenarios in an online survey.}
  \label{fig:methodology_overview}
  \vspace{-5mm}
\end{figure}

Figure~\ref{fig:methodology_overview} shows an overview of our employed analysis process. In this section, we give a bird's eye view on the process as a whole. Subsequent sections then describe in full detail how we realised each of the individual steps.
Overall, our research method involves six steps, structured into two larger phases of three steps each.

\subsection{Phase 1: Extracting User Problems}
In the first phase, the goal is to extract concrete problems that users face in everyday interactions with intelligent systems. This phase has three steps (see first row in Figure~\ref{fig:methodology_overview}):
\begin{enumerate}
    \item \textit{Collecting user experiences:} In this first step, we collect experiences that users had with intelligent systems. In particular, our approach aims to gather such experiences from ``natural'' sources that already exist, in contrast to assessing them with a user study. In our specific work here, we used app reviews on the Google Play Store. In general, other suitable sources of such reported experiences could involve personal blogs, social media posts or even internal information, such as support tickets, if accessible.
    \item \textit{Finding the main topics:} The second step filters and structures the collected user experiences from step one to identify the main topics. In this work, we used ratings and sentiment analysis to sort out irrelevant posts and employed statistical topic modelling to cluster the reviews by topic automatically. In general, this step could be realised quantitatively, as in our case, or qualitatively through manual analysis. This choice will likely be informed based on the expected semantic complexity of the aspects of interest, the dataset size, and other factors.
    \item \textit{Extracting user problems:} Finally, the third step of phase one extracts user problems from the collected and structured data. Depending on the respective research question, this step is likely to exclude part of the data. In our case, we manually inspected the ten most representative reviews associated with a topic as discovered by the model to then extract the underlying core problems. In general, we expect that such manual analysis is crucial in this step, since this process of problem extraction from a larger dataset necessarily involves semantic aggregation and abstraction.
    \setcounter{methodEnumCounter}{\value{enumi}}
\end{enumerate}
In summary, these three steps result in a set of concrete problems that users have faced with an intelligent system in the past, backed by existing empirical evidence provided by said users. As outlined above, in this work, we implement these three steps by analysing app reviews, both with statistical topic modelling and manual analysis. Section~\ref{sec:data_analysis} describes these steps in detail.

\subsection{Phase 2: Developing Solutions for Support}
In the second phase, the goal is to develop solution principles for the extracted problems, while again considering the views of real users. At the same time, this second phase validates the extracted user problems and allows to incorporate opinions from other users than in the first phase.
This phase also consists of three steps (see second row in Figure~\ref{fig:methodology_overview}):
\begin{enumerate}
    \setcounter{enumi}{\value{methodEnumCounter}}
    \item \textit{Representing user problems:} In this fourth step, the extracted user problems are represented in a way that can be shown to (other) users of the target group that shall be supported through the research outcome, for example, the developed solutions. In our work here, we represented each extracted problem with one written scenario (1-3 sentences). In general, alternative representations may also use images, videos, prototype systems/demonstrators, or live demonstrations, depending on the intended study design and setup.
    \item \textit{Asking users about the problems:} This fifth step uses the created problem representations to ask users about the corresponding problems. Here, we used an online survey that presented the written scenarios along with several open questions. More generally, this step could also employ qualitative research methods, such as interviews, think-aloud walkthroughs, journaling, and others. The specific questions may vary depending on the research goal. Nevertheless, Figure~\ref{fig:methodology_overview} shows the three questions we posed here, which we judge as fundamental enough to provide a good starting point for most studies interested in designing solutions.
    \item \textit{Informing solution principles:} This final step encompasses the processing and analysis of the gathered user responses from step five. What exactly this looks like depends heavily on the goal of the research overall, yet it fundamentally involves interpretation of the results with regard to design choices, future research, iterations of the method, and so on. For example, thematic analysis of the qualitative data can be used to derive a code book and subsequently code user coping and support strategies. %In our case, we discuss suggested support strategies and broader implications for designing user support in line with the assessed user views.
\end{enumerate}
In summary, these three steps serve the development of solutions for the extracted user problems, backed by empirically assessing the target user group's views, in particular on coping and support strategies. As outlined above, in this work, we realise these three steps by presenting problem scenarios in an online survey and aggregating users' coping strategies and wishes for support. Section~\ref{sec:online_survey} describes these steps in detail.

\section{Phase 1 -- Review Analysis}\label{sec:data_analysis}

In this section, we explain how we realised the first three steps of our research method in detail (cf. Figure~\ref{fig:methodology_overview}). As shown in the previous section, it is challenging to directly ask users about their experiences and problems with intelligent applications.  
Providing one solution, Bucher~\cite{Bucher2017} examined people's personal stories about the Facebook algorithm through tweets and interviews. We followed a similar approach but chose to analyse \textit{reviews} to increase the sample size. Reviews reflect users' attitude towards an application and document the experiences they make~\cite{guzman2014, maalej2016}. We scraped reviews from the Google Play Store and extracted and clustered user-reported problems both with machine learning and manual analysis. 

\subsection{Data Acquisition}
We built a web crawler to scrape the latest 10,000 reviews for \emph{Facebook}, \emph{Netflix}, and \textit{Google Assistant} respectively, as well as 15,448 reviews for \emph{Google Maps} from the US Google Play Store. 
Since we wanted to analyse reviews and problems in detail, we decided to take a \emph{snapshot} of current reviews instead of scraping millions of reviews over a longer time period. This sample size allowed us to include manual analysis and human interpretation in addition to quantitative data analysis. We reflect further on this choice in our discussion.
An overview of the dataset sizes throughout the analysis can be found in Table~\ref{tab:data-analysis}. 

\begin{table}
    \renewcommand{\arraystretch}{1.5}
    \setlength{\tabcolsep}{4pt}
    \centering
    \scriptsize
    \begin{tabularx}{\columnwidth}{p{2cm}p{3cm}XXXX}
        \toprule
         & Review Period & Scraped & $\leq$ 3 Stars & Sentiment & Topic \\
        \midrule
        \textbf{Facebook} & 15 Aug -- 19 Aug 2018 & 10,000 & 2,894 & 2,216 & 2,196\\
        \textbf{Netflix} & 23 Jul -- 19 Aug 2018 & 10,000 & 2,887 & 2,345 & 2,340 \\
        \textbf{Google Maps} & 30 Jun -- 19 Aug 2018 & 15,448 & 4,015 & 3,019 & 3,018\\
        \textbf{Google Assistant} & 7 April -- 19 Aug 2018 & 10,000 & 2,842 & 2,406 & 2,406\\
        \bottomrule
    \end{tabularx}
    \caption{Overview of dataset sizes throughout analysis.}
    \label{tab:data-analysis}
\end{table}

\subsection{Preprocessing}
In line with findings by Maalej et al.~\cite{maalej2016}, a first manual inspection of the data indicated that most reviews praised the app or simply repeated the star rating. Since these reviews were uninformative for our analysis of \emph{problems} with intelligent apps, we excluded all reviews with higher ratings than three stars (out of five stars). As a consequence, the dataset size was decreased by 71 to \SI{74}{\%} respectively (cf. Table~\ref{tab:data-analysis}). We also removed reviews written entirely in non-Roman script. 

To obtain a more nuanced impression of users' attitudes towards the apps, we conducted a sentiment analysis of the remaining reviews as recommended by Maalej et al.~\cite{maalej2016}. We used the Google Cloud Natural Language API\footnote{https://cloud.google.com/natural-language/} to assign each review a score and a magnitude value. The score indicates a positive or negative emotion of a review, while the magnitude refers to the amount of emotional content within a review~\cite{googleAPI2018}. For our analysis, we selected all reviews with negative emotional content (score $\leq$ $-$0.1). Since the magnitude is proportional to a text's length, we accepted all reviews with a magnitude $\geq$ 0.1. 
The final number of reviews for each app can be found in Table~\ref{tab:data-analysis}.

\subsection{Extracting Problems from Reviews}
We adopted a two-step data analysis approach with machine lear\-ning and manual analysis to identify user-reported problems in the remaining 9,960 reviews. This allowed us to combine the benefits of both data-driven analysis (e.g. covering a large dataset) and human interpretation (e.g. insight into subtle nuances of user problems).

\subsubsection{Statistical Topic Modelling}
In the first step, we applied topic modelling to break the data down into topics.
In particular, we used Mallet's (MAchine Learning for LanguagE Toolkit) implementation via Gensim~\cite{mccallum2002} of Blei et al.'s Latent Dirichlet Allocation (LDA) algorithm~\cite{blei2003}. We applied this modelling approach to each app's set of reviews separately. 

For this step, the reviews were further preprocessed by removing punctuation, superfluous characters (e.g. emojis), and stopwords (i.e. common words such as ``the''), and through lemmatization and tokenization. We incorportated bigram and trigram representations to allow the model to account for phrases that consist of more than one word (e.g. ``please bring back''). This removed reviews consisting of stopwords only.  

The main hyperparameter for the LDA model is the expected number of topics. To inform this, we tested 25 models for each application with topic numbers ranging from two to one hundred. We then chose the model with the highest coherence score. In cases without a peak coherence score, we selected the smallest number of topics that resulted in a score of at least 0.5 to keep the number of topics at a manageable size for our manual analysis step.

\subsubsection{Manual Analysis}
In the second step, we manually labelled the topics based on the top ten keywords as well as the ten most representative reviews associated with a topic as discovered by the model (1,040 reviews out of 9,960 in total). In other words, we manually extracted and coded the core \textit{problem} underlying each topic. This was done independently by two researchers, who then synthesised their analysis. 
Finally, we aggregated the resulting problems into more abstract \textit{problem categories} identified across all three apps. 

%% file: sections/04-results-topics.tex
\section{Results: User-Reported Problems}
Topic modelling yielded a total of 22 topics for Facebook, 34 topics each for Netflix and Google Maps, and 14 topics for Google Assistant. For 21 (Facebook), 33 (Netflix), 34 (Google Maps) and 14 (Google Assistant) of these topics, we extracted an underlying core problem. Reviews in the remaining two topics regarding Facebook and Netflix did not indicate a distinct underlying core problem and were therefore excluded for further analysis. Other topics related to bugs and usability issues were discarded as well. The remaining sample consisted of three topics each for Facebook and Google Assistant as well as eleven for Netflix and 13 for Google Maps. In total, these topics covered 12.5\% (Facebook, 274 reviews out of 2,196), 29.7\% (Netflix, 694 reviews out of 2,340), 35.5\% (Google Maps, 1,072 reviews out of 3,018) and 27.3\% (Google Assistant, 658 reviews out of 2,406) of the datasets used for topic modelling.
We aggregated our results into problem categories along up to four steps of a basic pipeline of algorithmic decision-making and related interactions: (1) knowledge base, (2) algorithm, (3) user choice, and (4) user feedback. \textit{Knowledge base} comprises problems related to the \textit{database} of the system. \textit{Algorithm} includes problems about the \textit{algorithmic decision-making}. \textit{User choice} refers to problems with options to \textit{control and influence} the algorithmic decision-making and to generally express and have a choice. Finally, \textit{user feedback} covers problems related to options for feedback and correction. Employing this approach, our analysis not only focuses on the algorithm itself, but also captures the interplay between algorithmic decision-making and users more broadly.

All identified problems, their assignment to the problem categories and exemplary reviews can be found in Table~\ref{table:results}. All user quotes have been reproduced with original spelling and emphasis.

\begin{table*}[htbp]
\input{sections/table_new.tex}

\label{table:results}
\end{table*}

\begin{table*}[htbp]
\input{sections/table_new_2.tex}
\caption{Results of the review analysis: User-reported problems related to intelligent everyday applications that we further aggregated to more abstract problem categories. The columns on the right show the number of reviews assigned to each problem category and an example review to illustrate the underlying problem (original spelling and emphasis).}
\label{table:results}
\end{table*}

\subsection{Facebook}

\subsubsection{Problems with Curation Algorithm}
The overall content curation of the Facebook news feed seemed biased to users and they accused Facebook of deliberately concealing posts. Furthermore, users reported problems with the content composition and ranking of the feed: They complained that they were not able to arrange the news feed in chronological order and that they missed posts by their friends and family. Instead, they had the feeling of being ``spammed with posts from pages [they] follow'' and advertisements.  

\subsubsection{Problems with User Choice}
Users expressed the desire for more control over %intelligent features of 
their feed. For example, they asked for the option to remove intelligent components, such as the \emph{People You May Know} section or to turn off \emph{Marketplace} notifications. 

\subsection{Netflix}

\subsubsection{Problems with Knowledge Base}
Users criticised Netflix' basis for recommendations, namely the available films and shows. First, they complained about the uninteresting selection of films in general (``[...] for a film lover myself, it's grown stale. [...]''). Furthermore, users claimed that the available films were outdated and their favourite films had been removed (``When you see that they have added new movies that you are interested in, make sure to watch them quickly because they'll probably remove it within a week.''). They also accused Netflix of a biased content selection by favouring their own original content over external content (``[...] now I've noticed that Netflix is starting to kind of put on a lot of their own stuff and they're taking out the movies and TV shows that I want to watch [...].''). Finally, users disapproved of incomplete TV shows for which not all seasons are available.  

\subsubsection{Problems with Recommender Algorithm}
Users complained about a mismatch between Netflix' recommendations and their interests (``You guys don't even give good recommendation anymore''). In particular, they noted that notifications about new releases and top picks were not to their taste (``Just got one [notification] telling me Better Call Saul 3 is out. I don't care and don't want that clutter in the way of my workflow''). 

\subsubsection{Problems with User Choice}
According to the reviews, users would like to have more options for selecting interesting content themselves. They claimed that Netflix' current interface makes manual selection difficult due to missing information about films and TV shows. For example, users requested a synopsis and description about an item since they ``[needed] more than a still photo and search keywords to decide if [they wanted] to watch something''. Moreover, the interface misled users and caused confusion about the language of an item: While all film titles were shown in their system's language, users did not know whether an item was actually available in this language until they started watching. Finally, the reviews indicated an incomplete search coverage since the search functionality only returned ``irrelevant results''.   

\subsubsection{Problems with User Feedback}
Users indicated that the binary rating system (thumbs up/down) Netflix currently uses is not sufficient to provide meaningful feedback, which in turn might be reflected in the relevance of recommendations. Users noted that the provided certainty of a recommendation \textit{always} seemed to be very high. Hence, a user asked: ``How am I supposed to know if I should actually watch the movie if every movie is a match''.  

\subsection{Google Maps}

\subsubsection{Problems with Knowledge Base}
The reviews indicated problems with the app's basis for routing suggestions. For example, users complained about inaccurate location sensors when Google Maps ``is unable to pin point [their] location''. In particular, users experienced these problems off-road. For example, a user described that her location ``appears to shift kilometres or more in seconds'' when she goes hiking. In addition, users encountered missing or incorrect map information when the application sent them ``off to a different location than you expect''.

\subsubsection{Problems with Routing Algorithm}
Users experienced several problems with the routing algorithm. First, the time estimate for a route seemed inaccurate and inconsistent to them. For example, one user recounted: ``Keeps adding completely random routes, as well as 10 minutes extra travel time to the same route that I take every day''. Other problems were unstable route selection and incorrect directions which led to users getting lost. Furthermore, users reported that Google Maps announced last minute turns and surprised them with sudden route changes or suggestions of dangerous and illegal manoeuvres such as a U-turn on the highway. Users also strongly criticised that the routing algorithm ignored their settings. Although they had explicitly excluded tolls, the app guided them via toll roads in the end.  

\subsubsection{Problems with User Choice}
Users experienced problems when they tried to override algorithmic decision-making by manually selecting a route or location. They indicated the need for more and more specific routing options to match their needs. For example, they recommended a mode for trucks or big cars which avoids narrow streets. Users would also appreciate alternative routing criteria, such as a scenic route. They further found that their manually selected routes were suddenly overwritten by Google Maps ``without telling'' them. Finally, users were frustrated when their knowledge was ignored but felt that their routing decision would have been better than Google Maps'. In particular, they complained that Google Maps repeatedly led them through a construction site. 

\subsubsection{Problems with User Feedback}
Users encountered problems when trying to give feedback or corrections to the app, such as missing or incorrect data points. They claimed that Google Maps ``has failed to respond to fix the flaw''. For example, a user indicated that Google Maps ``consistently thinks [her] home is on another continent''. Similarly, users reported that Google Maps did not allow them to easily report road construction sites. The reviews suggested that these reports were not taken into account ``even if reported by many users for 4 weeks''.

\subsection{Google Assistant}
For Google Assistant we only found \textit{problems with the Natural Language Processing algorithm}. First of all, users had the impression that the assistant could not understand their speech input, claiming that ``I speak clear English but it never gets what I ask correct.'' Furthermore, users complained that although the Assistant ``can transcribe the words which I said correctly, it never understands what I want to do.'' For example, a user wrote that the Assistant did not enter a doctor appointment in the calendar but instead showed the location of nearby doctors. On the other hand, users mentioned that they felt ignored by the Assistant. One user explained that the Assistant ``think[s] about [the command] for a bit and then do[es] absolutely nothing'', giving the impression of a ``surly shop assistant who can't be bothered to help''. Another user complained that the Assistant simply did not execute a reminder set earlier: ``Ok Google, remind me to take the pizza off the oven in 10 minutes. Pizza burnt.''

%% file: sections/table_new.tex
\newcolumntype{L}{>{\raggedright\arraybackslash}X}
\renewcommand{\arraystretch}{1.9}
\setlength{\tabcolsep}{4pt}
\centering
\scriptsize
%\begin{tabularx}{\textwidth}{ >{\raggedright\arraybackslash}p{2.0cm} >{\raggedright\arraybackslash}p{2.8cm} l X }
\begin{tabularx}{\textwidth}{  >{\raggedright\arraybackslash}p{1.75cm} >{\raggedright\arraybackslash}p{2cm} l X }
    \toprule
    \textbf{Problem Category} & \textbf{Problem} & \textbf{\# Rev.} & \textbf{Example Review} \\
    \midrule
    \multicolumn{4}{c}{\textit{Facebook}} \\
    \midrule
    \rowcolor{rowcol} Algorithm & Biased content curation &  117 & ``Cowards, censuring videos made by Republicans! [...] Why not let people post what they want?! I want to see Will Witt's videos, but you don't let me because you don't agree with him. [...]'' \\
    \rowcolor{rowcol} & Selective content composition and ranking & 81 & ``[...] You know what i miss, seeing actual post from friends. Not ads, not what my friends liked or commented on. What they post and share. Not spammed with post from pages I follow. Things ACTUAL FRIENDS POST. [...]''\\
    User choice &  Insufficient options for intelligent components & 76 & ``I get irritated with that People You May Know. I tried that I don't want to see this option but it doesn't work anymore. Please kindly give us any option to remove this thing.'' \\
    \midrule
    
    \multicolumn{4}{c}{\textit{Netflix}} \\
    \midrule
    \rowcolor{rowcol} Knowledge base & Uninteresting content & 81 & ``It used to be good, but now for a film lover myself, its grown stale. [...] Unfortunately, once you watch many of whats on there, it just gets bland and you wind up scrolling through things you've already been through. '' \\ 
    \rowcolor{rowcol} & Outdated content & 64 & ``[...] I want to say that Netflix is getting really old, that and it picks some of the worst movies I've ever seen especially when it comes to horror movies. [...]'' \\
    \rowcolor{rowcol} & Biased content selection & 44 & ``I used to love Netflix a lot cuz they had all the shows I wanted and movies but now I've noticed that Netflix is starting to kind of put on a lot of their own stuff and they're taking out the movies and TV shows that I want to watch [...].'' \\ 
    \rowcolor{rowcol} & Incomplete content & 74 & ``At first i thought its really worth to pay BUT when i'm on the end season of series, its incomplete [...].'' \\

    Algorithm & Mismatch between recommendations and user interest & 59 & ``You guys dont even give good recommendation anymore. [...] For example how in anyway is Godzilla 2 related to midnight run with Robert de niro'' \\
    & Mismatch between notifications and user interest & 39 & ``I'm still receiving notifications from the app. Just got one telling me Better Caul Saul 3 is out. I don't care and don't want that clutter in the way of my workflow'' \\
    \rowcolor{rowcol} User choice & Missing content description & 96 & ``They just removed the synopses for all their shows... What gives Netflix? I need more than a still photo and search keywords to decide if I want to watch something'' \\
    \rowcolor{rowcol} & Missing information & 26 & ``They've now removed all descriptions of the show or movie. [...] remember exactly what every single thing you ever put in your watchlist is about'' \\
    \rowcolor{rowcol} & Misleading information & 59 & ``Its just I speak English and it don't even tell you its in another language until you start watching it. Please add more English/American movies or at least state their not in English.'' \\
    \rowcolor{rowcol} & Incomplete search coverage & 94 & ``The search system is poor. A search returns with irrelevant results. Can you please make it basic search?'' \\
    User feedback & Insufficient feedback options & 59  & ``The rating system is still horrible, every movie I look at says 98\% match like how am I supposed to know if I should actually watch the movie if every movie is a match. Bring back the star system.'' \\
    \midrule
    
    \multicolumn{4}{c}{\textit{Google Maps}} \\    
    \midrule
    \rowcolor{rowcol} Knowledge base & Inaccurate user location & 206 & ``Maps is unable to pin point my current location.''\\
    \rowcolor{rowcol} & Inaccurate user location (off-road) & 85 &  ``This works well when you are on a road. However, take it off road, and its useless. Its laughable when I go hiking, and share my location with my wife. If I'm not on a road, my location appears to shift a kilometre or more in seconds: I even appear to walk on water, crossing lakes and rivers without any effort!'' \\
    \rowcolor{rowcol} & Incorrect map information & 79 &  ``BEWARE: This app could send you off to a different location than you expect. Check the address multiple times and make sure your address is correct first {[}...{]}'' \\

    \bottomrule
\end{tabularx}

%% file: sections/table_new_2.tex
\newcolumntype{L}{>{\raggedright\arraybackslash}X}
\renewcommand{\arraystretch}{1.9}
\setlength{\tabcolsep}{4pt}
\centering
\scriptsize
%\begin{tabularx}{\textwidth}{ >{\raggedright\arraybackslash}p{2.0cm} >{\raggedright\arraybackslash}p{2.8cm} l X }
\begin{tabularx}{\textwidth}{  >{\raggedright\arraybackslash}p{1.75cm} >{\raggedright\arraybackslash}p{2cm} l X }
    \toprule
    %\textbf{Problem Category} & \textbf{Problem} & \textbf{\# Rev.} & %\textbf{Example Review} \\
    %\midrule

    %\multicolumn{4}{c}{\textit{Google Maps}} \\    
    %\midrule
    %\rowcolor{rowcol} Knowledge base & Inaccurate user location & 206 & ``Maps is unable to pin point my current location.''\\
    %\rowcolor{rowcol} & Inaccurate user location (off-road) & 85 &  ``This works well when you are on a road. However, take it off road, and its useless. Its laughable when I go hiking, and share my location with my wife. If I'm not on a road, my location appears to shift a kilometre or more in seconds: I even appear to walk on water, crossing lakes and rivers without any effort!'' \\
    %\rowcolor{rowcol} & Incorrect map information & 79 &  ``BEWARE: This app could send you off to a different location than you expect. Check the address multiple times and make sure your address is correct first {[}...{]}'' \\
    
    Algorithm & Inaccurate time estimate & 80 &  ``{[}...{]} Keeps adding completely random routes, as well as 10 minutes extra travel time to the same route that I take every day. Also it screws up distances by several miles and adds about 15 minutes to known estimated travel times for zero reason. \\
     & Unstable route selection & 53 &  ``Even though we have told them of a major error on a heavily traveled tourist route numerous times, they refuse to listen. Trucks are getting ticketed and tourists are getting lost at a rate of at least 15-20 per hour. {[}...{]}'' \\
     & Incorrect directions & 81 &  ``{[}...{]} gives incorrect directions or dramatic unnecessary detours {[}...{]}'' \\
     & Unsuitable turn advice & 67 &  ``It will tell you it has a faster route mid drive and automatically change routes on you unless you hit an option on your phone which is not only illegal but dangerous!! {[}...{]} Then when I finally got closer to my destination it told me to go down a two way street then through a one way no entry section, if a car had come around the corner while I was trying to reverse and turn around I could have been killed!'' \\
     & Routing ignores user settings & 87 &  ``{[}...{]} A month after my trip I received a toll violation of \$26 for being in the carpool lane. I had tolls turned off on the app but I didn't see the option on Android Auto. [...] (by the way you owe me \$26)'' \\
     
    \rowcolor{rowcol} User choice & Lack of routing options & 65 &  ``{[}...{]} YOU NEED TO GIVE US THE OPTION TO CHOOSE TRUCK ROUTES! {[}...{]} do you have any idea how scary it is if you miss your turn and Google Maps tries to reroute you down a road that you cannot fit on {[}...{]}'' \\
    \rowcolor{rowcol} \multicolumn{1}{c}{} & Overwriting user-selected route & 82 &  ``It offers me a route and I choose it because I want to take the secnic route. Then, without telling me just puts me back on the quickest route. Which drives me insane - not everyone its trying to get places fast some of us like to see the world while do it.'' \\
    \rowcolor{rowcol} \multicolumn{1}{c}{} & Ignoring user knowledge & 59 &  ``For the road closures that need satnav help it goes out of its way to Always route through them no matter what even if it takes longer.'' \\
    
    User feedback & Ignoring user map corrections & 69 &  ``Consistently insists my home is on another continent. Google have failed to respond to requests to fix the flaw, indeed they wont even acknowledge the problem. If it cant get my address right, how can we know other addresses are right Well, actually there are many corrections I've suggested, but I'm ignored, even though I'm right and Maps is wrong. Untrustworthy!'' \\
     & Ignoring user route corrections & 59 &  ``Google maps also doesn't have a way to report raod construction closures [...].'' \\
    \midrule
    
    \multicolumn{4}{c}{\textit{Google Assistant}} \\
    \midrule
    \rowcolor{rowcol} Algorithm & Understanding user commands & 249 & ``I speak clear English but it never gets what I ask correct so responses are useless. It won't complete call or message requests properly, and cant identify many songs, which is ridiculous. [...] This SUCKS'' \\
    \rowcolor{rowcol} & Interpreting user commands & 246 & ``It actually seems to be losing abilities and getting dumber! I used to be able to do things like entering an opticians appointment by talking to it. Now, although it can transcribe the words which I said correctly, it never understands what I want to do. Instead it shows me the location of some nearby opticians [...]''\\
    \rowcolor{rowcol} & Ignoring user commands & 163 & ``Ok Google, remind me to take the pizza off the oven in 10 minutes. Pizza burnt.'' \\
    \bottomrule
\end{tabularx}

%% file: sections/05-online-survey.tex
\section{Phase 2 -- Online Survey}\label{sec:online_survey}

Next, we explain in detail how we realised the second phase of our research method (cf. bottom row in Figure~\ref{fig:methodology_overview}). As motivated in Section~\ref{sec:method}, the goal of this phase is to inform solution principles for the extracted problems, supported by the target user group's own views. Therefore, we conducted an online survey to extend our understanding of the extracted problems with user feedback beyond the Google Play Store reviews. This further allowed us to collect statements on these problems from people who did \textit{not} write an app review about it. Crucially, we used the survey to assess users' own suggestions for solutions to the extracted problems.

\subsection{Survey Design}
In the survey, we described each extracted problem as a scenario, similar to how they were expressed in the original reviews. For example, we illustrated the \emph{Google Maps} problem \emph{Overwriting user-selected route} in the following scenario:

\begin{quote}
    \emph{Tom and Paula go for an outing. They selected the most scenic route on Google Maps. While on the road, Google Maps automatically switches to the shortest route.}
\end{quote}

For each scenario, participants were asked to (1) indicate whether and how often they had encountered a similar situation (Likert scale), (2) describe how they coped with this situation (open question), and (3) describe how the app could support them in this situation (open question). Each participant was randomly assigned ten scenarios to keep the survey short, especially considering the open questions. If participants indicated that they had never encountered a scenario, they were directly redirected to the next one without further questions.

Finally, participants were asked to indicate their average usage of the four apps, their smartphone or tablet's operating systems, as well as demographic information. 

\subsection{Participants}
We distributed the questionnaire via university mailing lists and social media. It was completed by 287 participants. Before analysis, we checked and cleaned all collected data. One participant was excluded from the final dataset due to invalid answers. The remaining sample consisted of 286 participants (\SI{68.5}{\%} female; mean age 24.5 years, range 17-67 years). 175 participants used an Android-based smartphone, 112 participants an iPhone, and nine participants a Windows-based smartphone.
\SI{64.0}{\%} used Facebook at least weekly, \SI{48.6}{\%} used Netflix at least weekly, \SI{78.7}{\%} used Google Maps at least weekly, and \SI{11.2}{\%} used Google Assistant at least weekly. Participants had a chance to win \EUR{50} in cash. 

\subsection{Data Analysis}
The first two authors reviewed all qualitative data to derive a coding scheme for the two open questions, namely participants' coping strategies and their wishes for support. Codes were extracted independently and then discussed together to construct a code book.
A random sample of \SI{10}{\%} of the answers of each scenario used in the survey was then coded by both coders independently using the code book. The results were compared and discussed to eliminate any discrepancies until a consensus was reached. The dataset was then split between both coders for the final coding. 
Table~\ref{tab:categories-experience} shows the number of participants who were presented with at least one scenario in a problem category and those who indicated having experienced one of these scenarios at least once. The last column shows how often the problems within one problem category had been experienced as median (scale: once, monthly, weekly, daily).

\begin{table}
\newcolumntype{L}{>{\raggedright\arraybackslash}X}
\renewcommand{\arraystretch}{1.5}
\setlength{\tabcolsep}{5pt}
    \centering
    \scriptsize
%\begin{tabularx}{\columnwidth}{lllll}
\begin{tabularx}{\columnwidth}{XXXXX}
\toprule
\multicolumn{1}{l}{\textbf{}} & \multicolumn{1}{l}{\textbf{Problem Category}} & \multicolumn{1}{l}{\textbf{\begin{tabular}[l]{@{}l@{}}\# Presented\end{tabular}}} & \multicolumn{1}{l}{\textbf{\begin{tabular}[l]{@{}l@{}}\# Experienced\end{tabular}}} &
\multicolumn{1}{l}{\textbf{\begin{tabular}[l]{@{}l@{}}Median\end{tabular}}} \\ 
\midrule
\textbf{Facebook}             & Algorithm                            & 220                                                                                                       & 138                   
                        & weekly
\\
                              & User choice                              & 91                                                                                                        & 71 
                              & weekly 
                              \\
\textbf{Netflix}              & Knowledge base                        & 167                                                                                                       & 92     & monthly                                                                                                \\
\textbf{}                     & Algorithm                             & 87                                                                                                        & 44      & monthly                                                                                           \\
\textbf{}                     & User choice                              & 200                                                                                                       & 93      & monthly                                                                                               \\
\textbf{}                     & User feedback                       & 104                                                                                                       & 37           & monthly                                                                                          \\
\textbf{Google Maps}          & Knowledge base                        & 208                                                                                                       & 114  & monthly                                                                                                  \\
\textbf{}                     & Algorithm                             & 235                                                                                                       & 99  & once                                                                                                 \\
\textbf{}                     & User choice                              & 207                                                                                                       & 85     & once                                                                                                \\
\textbf{}                     & User feedback                         & 177                                                                                                       & 41 & once                                                                                                    \\
\textbf{Google Assistant}                     & Algorithm                             & 199                                                                                                       & 43  & once                                                                                                 \\

\bottomrule
\end{tabularx}
\caption{Number of participants who were presented with a scenario relating to a problem category of the respective apps. Out of those, number of participants who had experienced one of these scenarios at least once. The last co\-lumn shows the median of how often the problems within one problem category had been experienced (scale: once, monthly, weekly, daily).}
    \label{tab:categories-experience}
\end{table}

\begin{table}
\newcolumntype{L}{>{\raggedright\arraybackslash}X}
\renewcommand{\arraystretch}{1.5}
\centering
\scriptsize
\begin{tabularx}{\columnwidth}{llXll}
    \toprule
\multicolumn{2}{l}{\textbf{Algorithm}}           && \multicolumn{2}{l}{\textbf{User Choice}} \\
 \midrule
\multicolumn{4}{c}{\textit{Coping Strategies}}                                                   \\
Control content directly                    & 36 && Control content directly                & 15  \\
Actively search for content information     & 27 &&                                         &     \\
Check settings                              & 6  &&                                         &     \\
Influence algorithm indirectly              & 3  &&                                         &     \\
Search for explanations                     & 3  &&                                         &     \\
\multicolumn{4}{c}{\textit{Support Strategies}}                                                  \\
Options to control news feed           & 43 && Choice of intell. components   & 18  \\
Prioritise in a user-centered way           & 21 && Options to control news feed       & 12  \\
Default chronol. order of news feed    & 20 && Prioritise in user-centered way       & 5   \\
Show non-personalised content               & 13 &&                                         &     \\
Let users filter content, not algorithm & 13 &&                                         &     \\
Give explanations                           & 18 &&                                         &     \\
\bottomrule
\end{tabularx}
\caption{Facebook: Selected user-reported coping and support strategies for each problem category (with number of mentions).}
\vspace{-4mm}
\label{tab:facebook}
\end{table}

%% file: sections/06-results-survey.tex
\begin{table*}
\newcolumntype{L}{>{\raggedright\arraybackslash}X}
\renewcommand{\arraystretch}{1.5}
\setlength{\tabcolsep}{5pt}
\centering
\scriptsize

\begin{tabularx}{\textwidth}{LlLlLlLl}
    \toprule
    \multicolumn{2}{l}{\textbf{Knowledge Base}}       & \multicolumn{2}{l}{\textbf{Algorithm}}                       & \multicolumn{2}{l}{\textbf{User Choice}}             & \multicolumn{2}{l}{\textbf{User Feedback}}           \\
    \midrule
    \multicolumn{8}{c}{\textit{Coping Strategies}} \\
Search for content manually          & 19         & Search for content manually                     & 14         & Get recommendations somewhere else           & 44         & Get recommendations somewhere else & 6          \\
Get recommendations somewhere else   & 6          & Ignore recommendation                           & 10         & Search for content manually                  & 27         &                                    &            \\
Try to ``reset'' algorithm           & 3          & Change settings                                 & 3          &                                              &            &                                    &            \\
\multicolumn{8}{c}{\textit{Support Strategies}} \\
Make content selection transparent   & 23         & Allow users to give feedback   & 12         & Provide more information about items & 50         & Improve rating system              & 26         \\
Provide more information about items & 6          & Use more fine-grained input & 4          & Include user reviews                         & 10         & Include user reviews               & 6          \\
Let users teach the system            & 2          & Let users teach the system                       & 4          & More filtering options                       & 6          &                                    &           \\
\bottomrule
\end{tabularx}
\caption{Netflix: Selected user-reported coping and support strategies for each problem category (with number of mentions).}
\label{tab:netflix}
\end{table*}

\section{Results: Coping Strategies \& System Support}\label{sec:results_strategies}
Participants reported a great variety of coping strategies and ways in which they would like to be supported by the system. Within the scope of this paper, we restrict our account to those strategies which we deem most relevant in terms of our research questions and for HCI research in general.
In the following sections, participants' quotes have been translated to English where necessary or have been reproduced with original spelling and emphasis. Since our final sample size only includes participants who completed the survey, participant IDs may be higher than 286.

\subsection{Facebook (Table 4)}

\subsubsection{Addressing Problems with Curation Algorithm}\label{subsubsec:fb_curation}
\textit{Coping Stra\-tegies:} Participants tried to influence Facebook's news feed mostly by directly controlling which posts are shown: They disliked pages they followed, unfriended people, or marked advertisements as spam.
Many participants felt that the best way to not miss out on their friends' news was to visit their page directly instead of relying on the news feed. P246 stated that she ``got into the habit of vi\-siting profiles of people whose news [she's] interested in''.
A few participants tried to indirectly nudge the algorithm to show specific content. For example, they visited their friends' pages or liked their posts and wrote personal messages more often.
Some participants also indicated that they had changed their settings to influence the content of the news feed, or had searched online for explanations of the algorithm's workings.

\textit{System Support:} While many participants already used options to organise their news feed, more settings were their most prominent wish for support. For example, P521 would like to be able to configure the feed in a way that posts by friends are displayed at the very top.
This was also reflected in the claim that the algorithm should prioritise posts in a more user-centred way. P518 suggested that the frequency of chatting with people as well as the frequency of visiting their pages should be considered.
Interestingly, many participants would like the news feed to be ordered chronologically by default. For example, P148 claimed to ``just want a news feed in which all friends and subscriptions appear in chronological order''. Some participants wanted to be able to switch algorithmic curation on and off, or even stated that content should only be filtered by the users, not by an algorithm. 
Participants were also interested in \textit{non-}personalised content even ``contrary'' (P48) to their opinion, or demanded support via explanations to better comprehend why a particular post was shown. 

\subsubsection{Addressing Problems with User Choice}
\textit{Coping Strategies:} Problems with user choice were similarly addressed using available options to control the feed content.

\textit{System Support:} Participants called for choice regarding the (intelligent) features that are included in the news feed, such as the ``people you may know'' feature presented in one of the scenarios. Moreover, the demand for more options or settings to control the feed was repeated in this problem category, as was the claim that the algorithm should more strongly take into account users' needs and wishes.

\subsection{Netflix (Table 5)}

\begin{table*}
\newcolumntype{L}{>{\raggedright\arraybackslash}X}
\renewcommand{\arraystretch}{1.5}
\setlength{\tabcolsep}{5pt}
\centering
\scriptsize
    %\begin{tabularx}{\textwidth}{p{3.5cm}lp{3cm}lp{4cm}lp{3.5cm}l}
    \begin{tabularx}{\textwidth}{LlLlLlLl}
    \toprule
    \multicolumn{2}{l}{\textbf{Knowledge Base}} & \multicolumn{2}{l}{\textbf{Algorithm}} & \multicolumn{2}{l}{\textbf{User Choice}} & \multicolumn{2}{l}{\textbf{User Feedback}} \\
    \midrule
    \multicolumn{8}{c}{\textit{Coping Strategies}} \\
    Manually without Google Maps & 42 & Deal with uncertainty & 20 & Manually without Google Maps & 33 & Manually without Google Maps & 17 \\
    Reload & 21 & Manually with Google Maps & 17 & Manually with Google Maps & 24 & Manually with Google Maps & 7 \\
    Recalculation & 19 & Manually without Google Maps & 10 & Deal with uncertainty & 9 & Feedback to Google Maps & 5 \\
    \multicolumn{8}{c}{\textit{Support Strategies}} \\
    Allow users to give feedback & 13 & Adapt to users & 15 & More options for routing & 16 & More routing alternatives & 8 \\
    Provide information about uncertainty & 7 & More routing alternatives & 13 & Inform early about critical parts & 16 & Allow user to give feedback & 5 \\
     &  & Inform early about changes & 11 & Ask for permission before changing route & 8 &  & \\
    \bottomrule
    \end{tabularx}
    \caption{Google Maps: Selected user-reported coping and support strategies for each problem category (with number \mbox{of mentions).}}
\label{tab:google maps}
\end{table*}

\subsubsection{Addressing Problems with Knowledge Base}
\textit{Coping Strategies:} Participants experienced problems related to Netflix' know\-ledge base, such as the lack of interesting content. They then tried to search the database manually for relevant films and shows to watch or looked elsewhere for recommendations, often online. Others tried to discover new content by creating a new additional account, to ``discover new films without the pre-built algorithm'' (P253). 

\textit{System Support:} Participants suggested making content selection more transparent to users, for example by displaying newest releases as the very first film category in the interface or by adding release dates. Participants also wished for more detailed information about films and shows. Some even mentioned that the selection of content should be based more on users teaching the system about their preferences.

\subsubsection{Addressing Problems with Recommender Algorithm}
\textit{Co\-ping Strategies:} Recommendations that participants perceived to be irrelevant were often ignored in favour of manual search. Some participants also stated that they had changed their profile settings to influence the algorithm's suggestions.

\textit{System Support:} In this problem category, participants generally called for expressive feedback options. They suggested that the algorithm should take their preferences into account on a more fine-grained level, such as through more detailed film categories and genre types.
For example, P393 suggested implementing ``mood-dependent filtering''. Others stated that the system could explicitly support discovery with ``films that do \textit{not} match [a users'] preferences'' (P253).

\vspace{-1mm}
\subsubsection{Addressing Problems with User Choice}
\textit{Coping Strategies:} Selecting films and shows on Netflix was often accompanied by using other sources for recommendations -- mostly asking friends or searching on popular rating websites. Interestingly, only one participant explicitly stated that she had followed Netflix' matching score (P352). 

\textit{System Support:} The majority of people presented with this problem called for more detailed descriptions of films, shows, and actors as a basis for accepting a recommendation. For example, they asked for including reviews and ratings from other platforms, and for bringing back user reviews. Such reviews sometimes even seemed to be more relevant to users' choices than system recommendations.
For example, P374 said that she would also like to see ``films that match my films only by 50 percent, but have been high-rated (by other users)''. Similarly, P415 stated that the system ``should show me more shows that all people like [...], not only those that I will probably like''.

\subsubsection{Addressing Problems with User Feedback}\label{subsubsec:netflix_feedback}
\textit{Coping Strategies:} The current way of giving feedback in Netflix via ``thumbs up/down'' was often criticised by participants for not being fine-grained enough. For example, P224 said that ``the old rating system was better, the new one does not provide a neutral expression or a mediocre one, but you have to choose one of two options''. P99 even said that with the implementation of the new system, she ``just got annoyed. I have not rated any films since out of frustration''.

\textit{System Support:} Absent any alternatives, some participants used the current rating option (P118), but expressed a desire for general improvements to the system. Many called for bringing back the star rating that Netflix had once used in order to give more nuanced feedback.
Other suggestions included an option for users to write reviews themselves, or the possibility to answer a short questionnaire at the end of a film.

\subsection{Google Maps (Table 6)}

\subsubsection{Addressing Problems with Knowledge Base}\label{subsubsec:maps_knowledge}
\textit{Coping Strategies:} To cope with problems regarding the app's knowledge base, participants indicated that they tried to find a location or route manually without Google Maps, for example by asking pedestrians or getting their bearings in the surroundings. Some participants also tried to reload the app or restarted GPS to trigger a recalculation. 

\textit{System Support:} Regarding support strategies, most participants simply suggested improving location sensing. Moreover, parti\-cipants asked for an option to give feedback to the system. For example, P288 wrote: ``Since I knew where I am, it would have been helpful to simply tell the app, this is my position, trust me''. To deal with situations in which location accuracy cannot be gua\-ranteed, participants suggested that the app should ``inform about \mbox{these uncertainties''.}

\subsubsection{Addressing Problems with Routing Algorithm}
\textit{Coping Strategies:} To prevent problems with routing, users pointed out that they tried to deal with uncertainty such as ``always adding an additional~\SI{20}{\%} on Google Maps' travel time'' (P273). Furthermore, participants reported that they selected the route manually in Google Maps. For example, P495 wrote that she ``compared the different routes regarding their length in Google Maps'' when she thought that Google Maps suggested a detour. Other participants indicated that they abandoned the app when facing routing problems.
When confronted with unexpected routing suggestions, participants explained that they drove more carefully, but this also negatively influenced their feeling of the routing reliability.

\textit{System Support:} To overcome these problems, participants indicated that Google Maps should adapt to their needs. Among other solutions, they suggested that the algorithm should include a user's usual driving or walking speed and adjust travel time accordingly. Additionally, participants proposed that the app should provide more routing alternatives, allowing users to choose between them. Finally, participants emphasised that Google Maps should inform them about changes in the route early, so that users ``have the opportunity to react in time'' (P15).

\subsubsection{Addressing Problems with User Choice}
\textit{Coping Strategies:} Several people experienced difficulties with Google Maps when they wanted to manually choose a route and override the algorithm's suggestions. As a result, most participants abandoned the app and tried to find a route manually. Alternatively, participants described that they deactivated the routing function and simply used the map to manually select a route. 

\textit{System Support:} To overcome these problems, participants suggested more use-case specific options for routing, e.g. for different car/truck sizes or scenic versus shortest route. Participants also expressed the wish to be informed early about critical parts of the route, such as ``highlighting difficult parts of the road in advance'' (P466), allowing users to prepare themselves or to manually change the route. Finally, participants stressed that Google Maps should ``not simply re-route but ask for permission before changing the route'' (P17). Several participants also suggested preventing re-routing by default.

\subsubsection{Addressing Problems with User Feedback}
\textit{Coping Strategies:} Participants indicated that they had repeatedly tried to give feedback to Google, but that this feedback was not being taken into account. When having the impression that Google Maps' suggestions led to delays or detours, most participants ignored the suggested routing absent any options for correction. For example, they manually tried to avoid a construction site by following a diversion, or used the map to manually find the address they were looking for. 

\textit{System Support:} Most participants suggested faster updates of the database to incorporate users' feedback. Participants also expressed the desire for more routing alternatives they could use to manually select a different route, for instance to avoid construction sites. Moreover, participants stressed the need for feedback and correction options. For example, P352 suggested that users should be able to ``mark a construction site for them personally, which is detected by the system to provide an alternative route''. 

\begin{table}
\newcolumntype{L}{>{\raggedright\arraybackslash}X}
\renewcommand{\arraystretch}{1.5}
\centering
\scriptsize
%\begin{tabularx}{\columnwidth}{p{4cm}lp{3cm}l}
\begin{tabularx}{0.6\columnwidth}{Ll}
    \toprule
\multicolumn{2}{c}{\textbf{Algorithm}} \\
 \midrule
\multicolumn{2}{c}{\textit{Coping Strategies}}  \\
Try again                  & 18 \\
Adjust interaction behaviour              & 16 \\
Stopped using Assistant                  & 10 \\
Stopped using a particular feature   & 3 \\

Use other service                & 3 \\
\multicolumn{2}{c}{\textit{Support Strategies}}  \\
Improve speech recognition algorithm         & 14  \\
Improve natural language processing algorithm         & 13 \\
Involve the user in case of uncertainty         & 10 \\
Allow users to give feedback   & 5 \\
\bottomrule
\end{tabularx}
\caption{Google Assistant: Selected user-reported coping and support strategies for the identified problem category (with number of mentions).}
\vspace{-4mm}
\label{tab:assistant}
\end{table}

\subsection{Google Assistant (Table 7)}
\textit{Addressing Problems with Natural Language Processing Algorithm.} 
\textit{Coping Strategies:} 
When participants experienced problems with Google Assistant's natural language processing algorithm, most of them simply tried again and repeated their command. Furthermore, several participants reported having adjusted their interaction behaviour to make it easier for the assistant to understand them. For example, P525 explained that she spoke particularly ``slowly'' and P259 tried to speak very ``clearly''. If these attempts failed, participants tried to write the command instead of using speech input. Consequently, some participants expressed frustration with the Google Assistant and either stopped using a malfunctioning feature (e.g. using the Assistant to set reminders) or stopped using the Assistant altogether. Some participants also indicated that they switched to another service (e.g. a reminder app) instead of asking the Assistant.

\textit{System Support:} The majority of the affected participants simply called for improving either the speech recognition or the natural language processing algorithm. Moreover, several participants noted that the system should ``show suggestions'' (P516) and involve the user if uncertain about the input instead of executing a wrong command. Some participants also suggested that the Assistant should ask for user feedback after an interaction to improve the algorithm in the future.

%% file: sections/07-discussion.tex
\begin{figure}[t]
  \centering
  \includegraphics[width=\columnwidth]{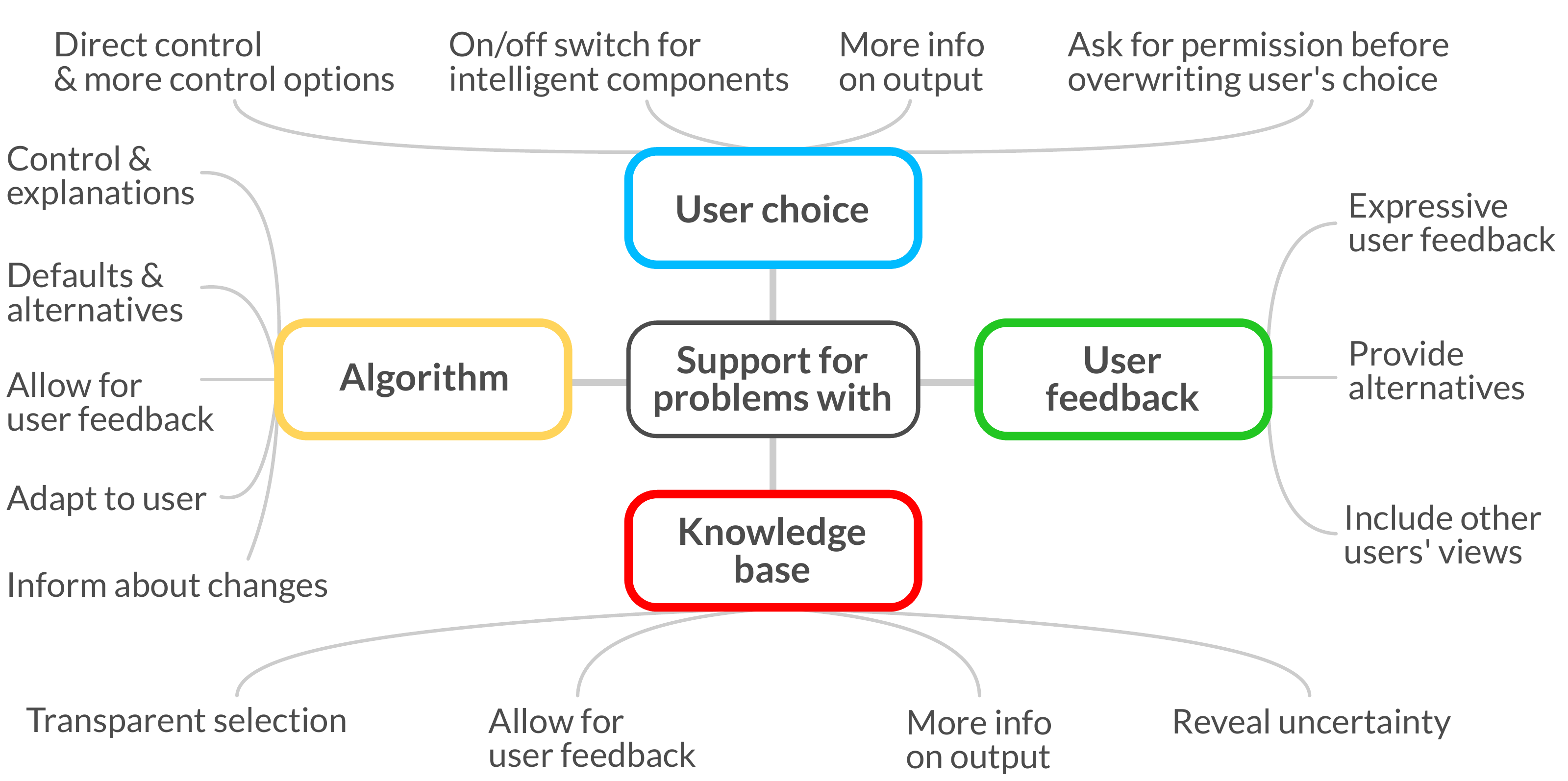}
  \caption{Users' suggested support strategies for their reported problems, aggregated across the studied apps. In general, we distinguish between support for problems with 1) what the system knows (red, bottom), 2) the algorithm (orange, left), 3) the opportunity for the user to express and/or have a choice (blue, top), and 4) the opportunity for the user to give feedback to the system (green, right). For instance, support for problems with the algorithm and user choice often focused on more (direct) control (e.g. of news feed content, Section~\ref{subsubsec:fb_curation}). As another example, users wished for feedback options on the system's knowledge (e.g. maps location, Section~\ref{subsubsec:maps_knowledge}), as well as for more \textit{expressive} feedback options (e.g. on movie recommendations, Section~\ref{subsubsec:netflix_feedback}). See Section~\ref{sec:results_strategies} for more such examples and details.}~\label{fig:overview_support}
  \vspace{-5mm}
\end{figure}

\section{Discussion}

\subsection{Implications for Design of User Support}
The identified problems and coping and support strategies revealed insights that span all 
four apps. Figure~\ref{fig:overview_support} provides a visual summary. Next, we discuss design implications 
using examples from the literature to illustrate how existing designs and prototypes could inspire practical solutions for the everyday systems we examined.

However, we acknowledge that these implications might increase the overall complexity of the interaction with the system. Future work needs to explore the trade-off between keeping interaction simple and still providing most of the functionalities requested by users.

\subsubsection{Let Users Stay in Control}
Control issues -- and the wish for more control -- reoccurred across all applications:
Users appreciate suggestions but want to make the final decision themselves. Systems should thus provide means for users to make such decisions on an informed basis: System suggestions should be presented with sufficient information to allow users to asses the value of a suggested item (e.g. film, route, post in news feed) themselves. 

Our analysis revealed that this information should primarily describe the suggested item \textit{itself} in an accurate way, instead of referring to how and why the suggestion was made. For example, participants called for more information about Netflix recommendations in order to assess their interest in a film or show rather than relying on the information about their personal matching score. Google Assistant users suggested that the system should actively involve them in case the system is uncertain about their input. In the context of intelligent everyday applications, this introduces a new perspective in contrast to the literature on explainability of intelligent systems, which so far has focused more on suppor\-ting users in understanding \textit{why} a system decision has been made (e.g.~\cite{Kay2013, Kulesza2015, Lim2011}).

Moreover, our results suggest that user control can be supported by \textit{revealing system uncertainty}. This might either refer to the system knowledge base or system \textit{input}, such as sensor or GPS data. In addition, such uncertainty is tightly coupled to the following algorithmic processing and thus also to uncertainty about system \textit{output}.

A recent example for revealing system \textit{input} uncertainty can be found in work by McCormack et al.~\cite{McCormack2019} in the context of musical improvisation with an intelligent system. In particular, the authors visualise the confidence of the system regarding its current joint play with the human musician. 

As an example of revealing \textit{output} uncertainty, Kocielnik et al.~\cite{Kocielnik2019} showed uncertainty in the decisions of an intelligent scheduling system via a simple chart, numbers, and different nuances of textual descriptions.

Similarly, everyday applications like Google Maps or Google Assistant could exploit different means of communicating uncertainty to the user. Google Maps might, for example, indicate uncertainty about the calculated time to reach a destination by displaying a time range. In particular, Google Assistant might harness the possibilities of spoken language and employ a richer vocabulary for indicating different levels of uncertainty  (e.g. "I'm not really sure...", "I'm certain that ...", "I'm positive that...", etc.).

Moreover, the system should not alter a decision that has been made by a user: Users want to have the last say. For example, both our analysis of app reviews and our survey revealed that people repeatedly complain about Google Maps automatically switching to a different route on the way.

\subsubsection{Provide more Fine-grained Control Options}
Our participants stressed the need for more fine-grained options to control algorithmic decision-making. While the analysis of Facebook revealed that many participants already have assumptions on how to make use of different settings to determine (part of) the content of their news feed, Google Maps users wanted to have more options to control route suggestions. For example, they would have liked to be able to indicate their vehicle type. Netflix users also called for more setting possibilities, such as to state their preferences for genres or mood of film.

An exemplary approach to meet the need for fine-grained user control has been presented by Kulesza et al.~\cite{Kulesza2015}. The authors introduce means for controlling the algorithm of an email spam classifier in the form of basic plots such as bar charts users can manipulate to influence and explore the system's workings~\cite{Kulesza2015}. This suggests what has been called \textit{meta-interaction}~\cite{Sarkar2015} in the literature, establishing an ``indirect'' channel between users and system. 

Moreover, Koch et al.~\cite{Koch2019} presented a system that supports design ideation in the form of mood boards co-created by system and users. Users can adjust system suggestions through widgets. These ``steering controls'' offer three levels of control, ``Not this one'', ``More like this'', and ``Surprise me''.

Similar options could be integrated into the everyday applications in focus of our work. Netflix, for example, already ``surprises'' users with personalised recommendations. It thus pushes content to users, instead of taking an information pull approach. Making this ``surprise'' option explicit might lead to a greater feeling of control on the user side. 

Another important theme emerging from our results was to make such options for control and explanation more obvious to users. For example, while Facebook already allows for a certain amount of control regarding items in the news feed, these options could be integrated in a more prominent way instead of hiding them in a menu.

In general, supporting these user needs likely requires the integration of control elements into the UI or reconsidering design and scope of already implemented elements of this type. 

\subsubsection{Explain Interactively} 
We found an overall need for explanation of system workings in the analysis of reviews and the online survey, even though it was not as prevalent as the desire for more control. Nevertheless, our results hint at the potential practical value of a more interactive approach to explanation, in contrast to the predominantly static one presented in the literature (e.g.~\cite{Rader2018, Ribeiro2016}): Our participants expressed a desire to \textit{try out} different settings and observe effects on the algorithmic output, possibly ``live''. For example, people suggested including different ``modes'' for the Facebook news feed, namely chronological order or algorithmically curated. This echoes a recent call by Abdul et al.~\cite{Abdul2018}, who suggest allowing users to ``[freely] explore the system's behaviour through interactive explanations'', for example, using interactive visualisations.

One interesting design direction is presented in work by Kapoor et al.~\cite{Kapoor2010} and Nguyen et al.~\cite{Nguyen2019}. Kapoor et al. allow users to directly manipulate the confusion matrix of a classification system to refine its decisions. Nguyen et al. explored interactive sliders for interaction with the underlying machine learning model: In this way, the features used in the model can be
manipulated. These sliders give users the possibility to observe the effects on the system's output and to adjust it accordingly. In particular, the authors provide an example UI for different machine learning models and also include an example movie recommendation system. % that uses Probabilistic Matrix Factorisation. 

While these solutions arguably target a level of technical involvement that might be undesirable for many users in everyday contexts, their approaches could inspire usable design solutions for expressive user feedback in everyday applications as well. For instance, Netflix users might benefit from an approach similar to Nguyen et al.'s movie recommendation example, with which they can influence and correct the system's suggestions.

\subsubsection{Allow Users to Turn Intelligence On and Off} 
We found that a large number of participants used the systems without their intelligent features: For example, users navigated themselves without Google Maps' intelligent algorithm, they tried to find film recommendations by asking friends or searching online, they used text instead of voice input to operate Google Assistant, or they directly visited friends' profiles they were interested in on Facebook. This behaviour was often used as a coping strategy for problems with the respective system, but could also be a feature of intelligent systems to stress user control or foster ``algorithmic awareness'' (cf.~\cite{Jhaver2018}).

This strategy can be found, for instance, in a meal recommendation system by Wasinger et al.~\cite{Wasinger2013}: Users receive a list of meals, recommended based on their personal eating preferences. The interface prominently includes a button that allows users to enable or disable this recommendation feature. In addition, visual highlighting informs users whether personalised recommendations are currently activated.

Following the many comments in our study asking for a simple chronological order of Facebook's news feed, such a functionality could also be applied to commercial products that rely on algorithmic content curation. This might not only raise awareness of algorithmic processing, but could also highlight this type of algorithmic curation as a distinct and useful feature that users feel in control of. On/off switches can also be seen as a kind of implicit explanation to explore which aspects of an application are influenced by an intelligent feature.

\subsubsection{Honour Expressive User Feedback and Corrections} 
Giving feedback and corrections to intelligent systems has been recognised as an integral part of interaction with such systems, in particular in the area of interactive machine learning~\cite{Amershi2014}. However, options for doing so in the practical systems we analysed in this paper were sparse or difficult to find for users, or were not seen as helpful in their current state. For example, Netflix offers a possibility for giving feedback, but the binary approach of ``thumbs up'' and ``thumbs down'' was heavily criticised both in the reviews and our online study. Users appreciate a more fine-grained, meaningful feedback system -- a 5-level star rating was often mentioned as a preferred approach. Google Assistant users even suggested that the system should not just offer the possibility, but ask for feedback \textit{proactively} to make sure that user feedback is incorporated to improve system performance.

Moreover, a seemingly obvious, but crucial follow-up issue is to actually \textit{take feedback into account} and provide a means of confirming this to the user.

\subsection{When in Doubt, Trust in Humans Seems to Outweigh AI}
Although our work did not focus on trust in intelligent systems in particular, our results indicate a tendency to favour human over algorithmic decision-making, not only in sensitive domains but also in comparably low-risk intelligent everyday applications. Decisions for films and shows on Netflix were mostly accompanied by drawing on \textit{human} recommendations elsewhere, either on rating platforms online or by asking friends directly. 
Moreover, some users of Google Maps estimated the arrival time themselves: From their experience, they had found that their own estimates better matched the time they actually required. These findings suggest that intelligent everyday applications are used as support for dealing with and organising daily issues and thus for \textit{augmenting} users' abilities, but not \textit{replacing} or readily exceeding them. This picture might be inverted for intelligent systems with abilities that humans lack, such as high-speed reaction times (e.g. self-driving cars), introdu\-cing randomness (e.g. a drawing program) or performing repetitive tasks (e.g. industrial robotics), which might present an interesting issue to explore in future work.

\subsection{Locating and Motivating Designs for User Support}
Reflecting on the identified problems, we highlight that they occur at different stages of algorithmic decision-making:
First, there are problems with \textit{input} and \textit{knowledge base} of said systems (e.g. GPS data, available maps and films).
Second, problems may occur related to the \textit{inference} process (e.g. content prioritisation, natural language processing, recommendation).
Third, problems can appear when the system presents or acts according to the \textit{output} of said inference (e.g. missing content and information, overwriting user choice).  
On a high level, these stages and associated problems indicate \textit{points of action} with which researchers and practitioners may explicitly frame and motivate a specific support concept or system. For example, a UI concept that adds a ``revise'' button next to algorithmic decision outputs could be explicitly motivated as addressing the output stage and the corresponding set of identified practical problems. In this way, a researcher or practitioner working on the concept could link her proposed solution to a problem category and specific exemplary problems which occur in practice.

\section{Reflection on Methodology}
We combined sentiment analysis and topic modelling with manual analysis to extract and aggregate user-reported problems. A similar combination of topic modelling and manual refinement has been used in recent data-driven literature surveys (e.g. see~\cite{Abdul2018}). This approach allowed us to explore a large dataset in a structured way (via topic modelling) and still look into user problems in detail via human interpretation. We believe that this approach is valuable for analysing experiences with intelligent systems beyond this paper: For example, it could also prove useful to extract reported problems on social networks, blogs, or services like Twitter.

While our approach allowed us to collect a variety of problems from a big sample of users, this collection might not be complete. Users who write reviews for the Google Play Store might not be a representative sample of all people using an application. Furthermore, we  excluded all reviews of users who were clearly frustrated with the application but did not describe a particular problem (e.g., ``shit app''). To overcome this problem, we included a broader sample of users in our online survey to validate the collected problems. Yet, this sample was also biased towards young users. Hence, we suggest that future work should focus on examining further problems, in particular with regards to user characteristics. For example, older users might suffer from different problems or experience problems in a different way.  

As a basis for our analyses, we took a \textit{snapshot} dataset of recent reviews. This could be repeated at a later point in time, or in regular intervals, to see how problems develop. A short-term sample might be influenced by events such as recent update releases. While our dataset contains reviews that relate to such events, the extracted main problems are clearly more long-lived and tied to fundamental app features and characteristics. We thus conclude that our snapshot was not overshadowed by a particular update. On the other hand, we cannot assess if we missed certain problems, especially rare ones. Nevertheless, any reoccurring dominant problem should also be reflected in our snapshot. 

In the online survey, we presented participants with specific scenarios since generally asking users about their problems with intelligent systems is challenging. This approach allowed us to gain a clear understanding of which problems users experience. Although we instructed participants to indicate whether they have experienced this particular problem or a similar one, it might have been difficult for participants to abstract the given problems and generalise them. For example, a participant might not have experienced a problem with selecting the most scenic route on Google Maps but a similar situation in which Google Maps overrode the user's selection. Thus, future work could qualitatively try to investigate the appropriateness of the scenarios and whether users are able to transfer them to similar situations they experienced.

In future work, a long-term dataset could be collected and ana\-lysed as well, potentially also covering further applications. However, it might then become more difficult to analyse the problems in nuanced detail, since manual analysis with human interpretation is challenging to scale up to millions of reviews. To address this, future analyses could use our work combining topic modelling and manual interpretation as a starting point (e.g. by training a problem classifier model on our extracted problems and \mbox{corresponding reviews}). Moreover, it would be interesting to assess whether the observed problems also hold true for other, non-consumer intelligent systems, e.g. research applications or complex expert systems. This comparison might allow to examine if problems mainly result from the system intelligence itself or are possibly influenced by other factors, such as conflicting needs of users and companies developing the applications.

Note that our work is based on users' experiences with the apps and their underlying intelligent algorithms. As our work was conducted without the involvement of Facebook, Netflix and Google, we have no insight into the details of the respective algorithm's workings at the time of data collection and analysis. Thus, we cannot verify if the problems that users reported can be attributed to actual problems in the respective algorithm. Different experiences might also be caused by different versions of the apps or algorithm. However, to inform support for users (e.g. explanations) it is the very problems that users \textit{experience} that are of interest to us, be they caused by actual malfunction of the system or not.

%% file: sections/08-conclusion.tex
\section{Conclusion}
The complex nature of intelligent systems with algorithmic decision-making poses difficult challenges for human-computer interaction (HCI) and often violates established usability principles, such as easy error correction and predictable output~\cite{Amershi2014, Dudley2018}.
This motivates work on supporting users during interaction, for example through explanations~\cite{Rader2018}. 

However, there is still little empirical evidence on practical everyday problems which people face when using intelligent applications. As a result, it often remains difficult for researchers to clearly link prototype solutions to empirically well-founded practical \mbox{user problems}.

To address this, we assessed and revealed such problems, as reported by users, by analysing 45,448 public app reviews with sentiment analysis and topic modelling.
We then enriched this data with information about users' coping strategies and support ideas through a follow-up online survey (N=286). Based on this information, we extracted problem types and corresponding implications for designing user support. 

In particular, our implications point towards supporting user control, feedback, and corrections, and reconsidering how and what to explain.
For example, our results suggest that explanations for recommendations should mainly describe the recommended content itself -- the how and why of the recommendation process is less interesting to users in our studied everyday practical context. This stands in contrast to the existing focus in the literature (e.g.~\cite{Kay2013, Kulesza2015, Lim2011}).

More generally, our work thus contributes empirical evidence for problem situations and facilitates a closer understanding of practical user problems in intelligent everyday applications.

\section{Project Resources}
The reviews with annotated problems and categories as well as the coded survey results are available via the project website: \url{http://www.medien.ifi.lmu.de/userproblems/}

\section{Acknowledgements}
We greatly thank Martin Eberl and the anonymous reviewers for their helpful comments on earlier versions of this article.
This project is funded by the Bavarian State Ministry of Science and the Arts in the framework of the Centre Digitisation.Bavaria (ZD.B).